\documentstyle[prl,aps,floats,twocolumn,graphics,epsfig]{revtex}

\title{Scale Invariance in disordered systems:
the example of the Random Field Ising Model}
\author{ Giorgio Parisi$^{1}$ and Nicolas Sourlas$^{2}$   }

\address{
 $^{1}$ Dipartimento di Fisica, INFM, SMC and INFN,
Universit\`a di Roma {\em La Sapienza}, P. A. Moro 2, 00185 Rome, Italy. \\
 $^{2}$ Laboratoire de Physique Th\'eorique de l' Ecole Normale
Sup\'erieure, 24 rue Lhomond, 75231 Paris CEDEX 05, France
 \footnote {*} {Unit\'e Propre du Centre National de la Recherche Scientifique,
associ\'ee \`a l' Ecole Normale Sup\'erieure et \`a l'Universit\'e de
Paris-Sud. }  \\ }

\date{\today}
\draft

\begin{document}
\twocolumn[ \hsize\textwidth\columnwidth\hsize\csname
@twocolumnfalse\endcsname
\maketitle

\begin{abstract}

We show by numerical simulations that the correlation function  
of the random field Ising model (RFIM) in the critical region 
in three dimensions has very strong fluctuations and that in a finite volume the 
correlation length is not self-averaging. 
This is due to the formation of a bound state in the underlying 
field theory. We argue that this non perturbative phenomenon is 
not particular to the RFIM in 3-d. It is generic for disordered 
systems in two dimensions and may also happen in other three 
dimensional disordered systems. 
 \end{abstract}


]

Ferromagnetic systems in a random fields (RFIM) have a puzzling behaviour.  
It is already known that 
perturbative renormalization group (PRG) leads to the prediction of dimensional reduction that is 
known to give  wrong results for the RFIM, and this although 
RFIM is, together with branched polymers\cite{ps2}, the only case where PRG can be analyzed to all 
orders of perturbation theory\cite{young,ps1}.  Several attempts were made to explain this 
discrepancy\cite{PARISIA,mezyoung,mezmon,brezdedo1,brezdedo2}.

One suggested possibility is the appearance of a non perturbative phenomenon (i.e. the formation of a 
bound state in the underlying field theory) which invalidates the perturbative analysis.  There are 
several consequences of this.  It is not any more obvious that naive scaling arguments apply in the 
scaling region as it is usually assumed, or if and how they should be modified.  We expect also 
these non perturbative phenomena, related to the formation of bound states, to invalidate the 
predictions of perturbative renormalization group for the critical behaviour of disordered systems 
(calculation of critical exponents, proof of universality classes etc.)  A new examination of these 
questions is required.

We claim that the appearance of bound states is not peculiar to the 3-dimensional RFIM 
(what is peculiar to the RFIM is the possibility to perform the very large scale simulations 
needed in order to observe these non perturbative phenomena). 
Kardar et 
al.\cite{kardar} have already noticed that averaging over the disorder generates attractive 
interactions among replicas that produce bound states in several two dimensional systems.  In the 
formation of bound states there is competition between the strength of the attractive forces and the 
size of the available phase space.  In two dimensions phase space is small and even a very small 
attraction wins, so bound states are generic.  In three dimensions formation of bound states depends 
on the strength of the attraction.  In the case of the RFIM Br\'ezin and De 
Dominicis\cite{brezdedo2} have already found that an instability appears in the kernel of the 
Bethe-Salpeter equation, leading to the formation of a bound state in less than 4 dimensions.  Our 
results confirm their analysis.  

The relevance of the attractive interaction among replicas can 
also be seen in a case where the interaction between replicas is 
repulsive.  This is the case of diluted branched polymers (or lattice animals).  There are not bound 
states in that case and perturbative renormalization group, i.e. dimensional reduction, has to work.  
Indeed dimensional reduction has recently been proven rigorously in this case\cite{bry}.

In this paper we address the question of the existence of bound states and of the validity of the usual 
scaling laws by studying the 3-d RFIM by extensive numerical simulations at zero temperature.  As 
many previous authors\cite{angles,ogielski,swift,ansou,sou,middlefish} we have taken advantage of 
its equivalence at zero temperature with the maximum flow problem in a graph\cite{anprram}, for 
which a very fast polynomial algorithm is known\cite{gold}.  The simulations mentioned 
above provided evidence of an infinite correlation length at the phase transition through a finite 
size scaling analysis, but no direct measurement of a correlation length was performed.  In the 
present paper we perform direct measurements of correlation lengths. We will see that the 
correlation functions have very strong fluctuations. We will show by a scaling analysis that 
these fluctuations are so important,  that in any  finite volume the 
correlation length in the critical region is not self-averaging (i.e. it is strongly sample dependent). 

Let us remind that the Hamiltonian of the RFIM is of the form 
$ H \  = \ -J \sum_{<i,j>} \ \sigma_{i} \sigma_{j} \ - \  \sum_{i} \
\ h_{i} \sigma_{i}   $. 
As usually $ \sum_{<i,j>} $ runs over neighbouring sites of a cubic lattice with coordinates 
$x,y,z=1,\cdots,L$ and $ h_{i} $ are independent random Gaussian variables with variance $ 
{\overline h_{i}^2 } = 1$.  It is well known that there is a zero temperature phase transition in 
this model.  For $ J > J_c $ the spins are ferromagnetically ordered, while for $ J < J_c $ we get a 
disordered phase where a large number of spins are locally aligned with the random external field.

In order to measure correlation length at zero temperature we proceed as follows.  We consider cubic 
lattices with linear size $L$.  For every sample of the random field, we considered two copies of 
the system with different boundary conditions.  For the first copy $\sigma $ all the spins at the 
plane $x=1$ are set to plus one.  For the other copy $\tau $ they are set to minus one.  We let free 
boundary conditions for both copies at the other end of the lattice, i.e. $x=L$.  We impose periodic 
boundary conditions on the two perpendicular directions.  For each value of the ferromagnetic 
coupling $J$ we find the two ground states $\sigma_i $ and $\tau_i $.  For every sample $s$ of the 
random field we measure the distance $d_s(x)={ 1 \over 2 } (1-q_s(x))$ where the overlap is given by
\begin{equation} \label{qs}
 q_s(x) = { 1 \over L^2 } \sum_{y,z} \sigma (x,y,z) \tau (x,y,z) 
\end{equation}
which depends on the sample $s$; (we remind that because of the boundary conditions at $ x= 1$, $ 
\sigma (1,y_0,z_0) \tau (1,y_0,z_0)= -1 $ ) $d_s(x)$ measures the proportion of spins which have 
opposite values in the two ground states at distance $x$ from the boundary.  By construction $d_s(1) 
= 1$.  If $J << J_c $ $d_s(x)$ will rapidly decrease with $x$, while for $J >> J_c $ it will 
asymptotically remain constant.
\begin{figure}[t]
\centerline{\hbox{\epsfig{figure=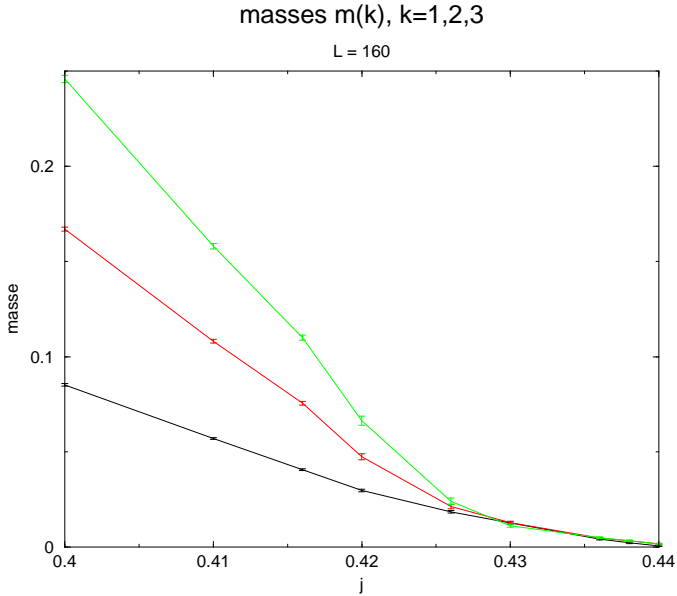,angle=-90,width=9cm}}}
\caption{Masses $m(k)^{1 / \nu } $ (with $\nu = 1.5 $ ) as a function of
$J$, for $L=160$. $k=1 $ is the lowest curve, while $k=3 $ is the highest. }
\label{fig_size}
\end{figure}
It is known from lattice gauge theories\cite{mapa} 
that one efficient way to measure correlation lengths, 
 is to impose boundary conditions on an observable and measure the 
variation of the observable with the distance from the boundary. 
From the behaviour of $d_s(x)$ measured numerically we are going to measure the 
correlation length and 
draw conclusions on the underlying field theory. 

We simulated 1200 samples for $ L = 160 $, 2000 samples for $ L = 120 $ and 3500 samples for $ L = 
80 $ for different values of the ferromagnetic coupling $ J \le J_c $.  From the previous 
simulations\cite{ogielski,swift,ansou,sou,middlefish} it is known that $ J_c \sim .44 $.  We 
found very strong sample to sample fluctuations of $d_s(x)$, even for systems of sizes $160^3$.  In 
order to extrapolate to the infinite volume limit, we studied the behaviour of $ d^{(k)}(x) = 
\overline{ d_s(x)^k } $ for $ k=1,2,3 $,  (as usually the bar denotes average over the random fields), 
i.e. the average of the $k$'th power of $d_s(x)$ over the samples.

For $ J < J_c $  we expect an exponential decay in $x$ of $ d^{(k)}(x) 
\sim \exp{( -m(k) x) } $. The masses $ m(k) $ are inverse correlation lengths. 
We first fitted our data with the formula 
\begin{equation} 
d^{(k)}(x) \ = \ { \exp{( -m(k) x )} \over x^{\alpha(k) } }  (a(k) + { b(k) \over x } )
\end{equation}
We expect $m(1)$ to vanish at the critical value $ J = J_c $.  According to finite size scaling $ 
m(1) \sim (J_c-J)^{\nu} f((J_c-J)L^{1 / \nu } ) $.  We found that the scaling function $f$ is a 
constant with a good approximation, i.e. a very small $L$ dependance of $ m(1) $, and $ 1.3 \le \nu 
\le 1.9 $, compatible with previous results\cite{ansou,sou,middlefish}.  We have obtained a better 
determination of $\nu $ in another set of simulations, (with different boundary conditions) that we 
will publish elsewhere.  For a given value of $J_c$ the statistical errors in $\nu $ are very small.  
The large uncertainty is due to subdominant corrections and to the extreme dependance of $\nu $ on 
the value of $J_c$.  For $ J << J_c $ (in practice for $J < .43 $) 
we found $ \alpha(1) = 0 $ in agreement with dimensional 
arguments, while for $J \sim J_c$ $ \alpha(1) = .27 \pm .05 $.
\begin{figure}[t]
\centerline{\hbox{\epsfig{figure=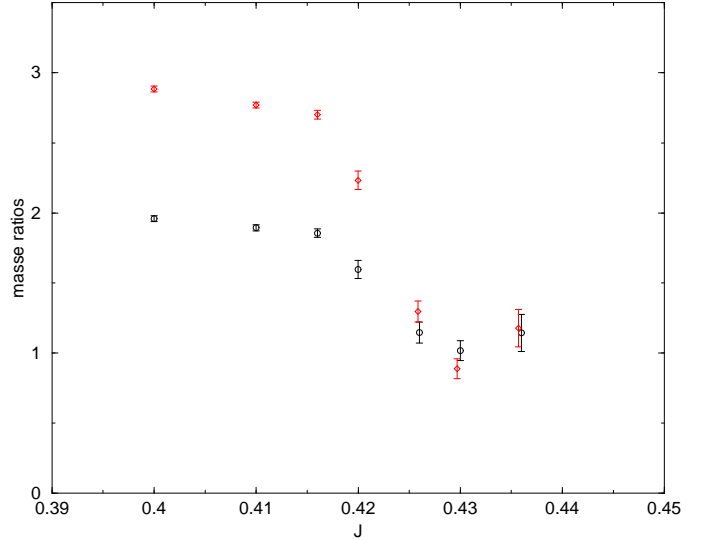,angle=-90,width=9cm}}}
\caption{Masse-ratios as a function of
$J$, for $L=160$. Circles represent $m(2)/ m(1) $ and
squares represent $m(3)/ m(1) $  }
\label{fig_size}
\end{figure}

In figure 1 we plot $ m(k)^{1 / \nu } $ (with $\nu = 1.5 $ ) as a function of $J$ for $L=160$. 
Not too far from $J_c$,  $ m(1)^{1 / \nu } $ 
is linear in $ J $ to a very good approximation, $ m(2)^{1 / \nu } $ and $ m(3)^{1 
/ \nu } $ have a more complex behaviour.  What is very surprizing is that for $ J > .425 $, $ m(2)$ 
and $ m(3) $ collapse with $m(1)$.  This is best illustrated in figure 2, where we plot the ratios 
$m(2)/m(1)$ and $m(3)/m(1)$ as function of $J$ for $L=160$.  Almost identical results were obtained 
for $L=120$ and $L=80$.  For small enough ferromagnetic coupling $m(k)/m(1)= k $, i.e. the masses 
are self-averaging, as expected in perturbative field theory\cite{sou0}, while close to the critical point 
$m(k)/m(1)= 1 $.  It is obvious that $ 0 \le d_s (x) \le 1 $ from the very definition of $d_s(x)$.  
It follows that $ d^{(k)} (x) \le d^{(1)}(x) $ and therefore $ m(k) \ge m(1) $; $ m(k) = m(1) $ 
amounts to the maximum possible violation of self-averaging of $ d_s (x) $.
\begin{figure}[t]
\centerline{\hbox{\epsfig{figure=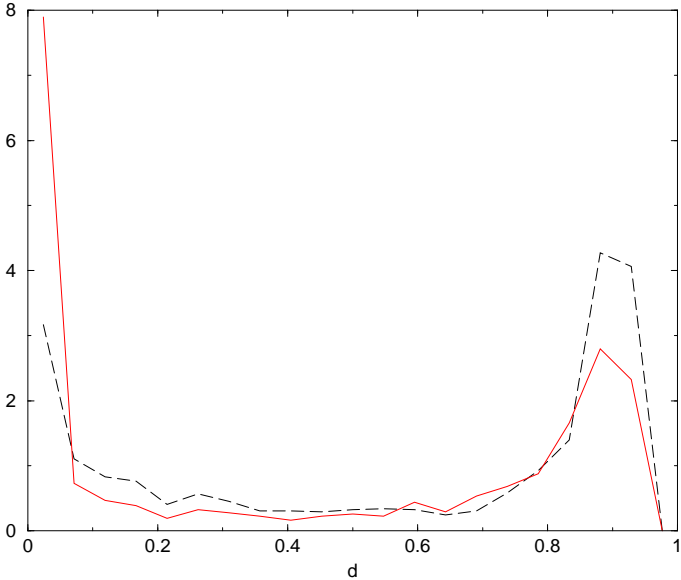,angle=-90,width=9cm}}}
\caption{ Probability distribution (histogramme)
of $d(x)$ for $J=.436$, $L=160$ and two values of $x$, $x=55$ (dashed line) and $x=110$ 
 (continuous line).  }
\label{fig_size}
\end{figure}
This is also illustrated in figure 3, where we plot the probability distribution (histogramme) $H(d)$ of 
$d_s(x)$ over the random field samples, for $L=160$, $J=.436$ and 
and two values of $x$, $x=55$ (dashed line) and $x=110$ 
(continuous line).  Theses histogrammes  
separate into two peaks, one around $ d=0 $ and the other around $ d=1 $.  As we vary $ x $ the shape 
of the histogramme remains the same, only the relative height of the two peaks changes.  For smaller 
but not too small $J$'s, in the region of $J$ where still $ m(k) \sim m(1) $,
 $ H(d) $ has also contributions 
outside the two peaks.  The peak around $ d=1 $ is necessary for $ m(k) = m(1) $, otherwise $ d^{(k)} 
(x) $ would decay with $x$ faster than $ d^{(1)} (x) $.

For any finite $x$ there is also the possibility of a superposition of exponentials in $ d^{(k)} (x) 
$.  We have therefore tried another fit to $d^{(2)} (x) $, 
\begin{equation} \label{d2}
d^{(2)}(x) \ = \ c_1 d^{(1)}(x)  +   c_2 (d^{(1)}(x))^2 +   c_3 \exp ( -m^{'} x )
\end{equation}
where $ m^{'} $ is a new masse. 
It turned out, particularly for intermediate values of $J$, that this is a much 
better fit. We found that no superposition of 
exponentials is needed in the fit of
 $d^{(1)} (x) $. We found that when $c_3$ is $ \neq 0 $, 
$ m^{'}  > 2 m(1)$, i.e. this term contributes only for small $x$. 

$d^{(1)} (x) $ has the slowest decay in  $x$ in equation \ref{d2}. Therefore 
$ c_1 $  measures the violation of self-averaging of $d$ at large $x$.
For small $J$, $ c_1 \sim c_3 \sim 0 $ i.e. the masse is self-averaging, while 
for $J \sim J_c $, $ c_2 \sim c_3 \sim 0 $, $ c_1 \sim 1$. 
In order to study the crossover 
between the two regimes, we tried the finite size scaling ansatz 
\begin{equation}
c_1 \ = \ g((J_c - J) L^{ 1 / \nu } )  \ = \ {\cal G } ( L m  )
\end{equation}
where $m=m(1)$. In the previous equation we used the finite size scaling of $m$, 
i.e. $m L = f((J_c - J) L^{ 1 / \nu } )$ to change variables from $ (J_c - J) L^{ 1 / \nu } $ to $m L$.
As is seen in figure 3 where $ c_1 $ is plotted versus $mL$
this ansatz works rather well. 
In the scaling region, i.e. for small $m$, 
$ c_1 \sim 1 $ and $ d^{(2)}(x) \sim \exp -mx $,
 i.e. $ m(2) = m(1) $ and we observe the maximum possible violation of 
self-averaging, as we have seen before. 
Outside the scaling region (i.e. for $m > .075 $)
we found that for constant $ J$, 
 $  c_1 \sim 1 / L^2 $ (we remind that in this region m depends on
$J$ and not on $L$ ) and $ d^{(2)}(x) \sim \exp -2mx $.
This means that  the violation of self-averaging of $d$ vanishes as $ 1 / L^2 $.
From the very definition of $d_s(x) $, 
 if $ d_s \ne 0 $, $ d_s \ge 1 / L^2 $. It follows that
$ d^{(2)}(x) \ge d(x)/L^2 $, i.e. $  c_1 \sim 1 / L^2 $ amounts to the 
smallest possible violation of self-averaging. 
It is, we think, very remarkable that in the scaling region we cross 
from the smallest possible violation of self-averaging, i.e. $  c_1 \sim 1 / L^2 $, 
 over to the maximum possible violation of self-averaging, i.e. $  c_1 \sim 1  $.
\begin{figure}[t]
\centerline{\hbox{\epsfig{figure=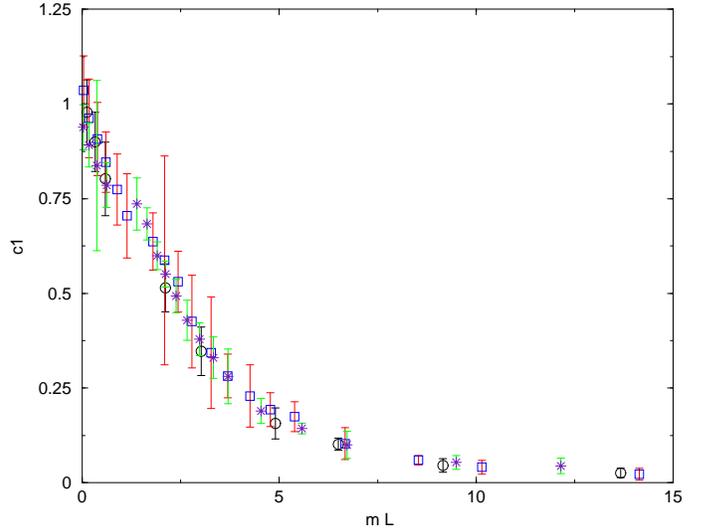,angle=-90,width=9cm}}}
\caption{ Scaling plot of $c1$ versus $mL$ (see text). Circles correspond to
$L=160$, squares to  $L=120$, and stars to $L=80$.}
\label{fig_size}
\end{figure}

 The quantities $d^{(k)} $ have a simple correspondence  
in field theory at finite temperature $T$.
Let us consider an infinite system at finite temperature.
We study the field theoretic equivalent to the quantity $q{_s}(x)$ defined in equation 
\ref{qs} and its correlations.
To the Ising spins $\sigma $ and $ \tau $ 
correspond the fields $\phi (\vec{x} ) $ and $\psi (\vec{x} ) $. 
Their vacuum expectation (i.e. their space average) 
 has been substracted from $\phi (\vec{x} ) $ and $\psi (\vec{x} ) $ so that their  
correlation, for example $ \langle \phi (\vec{x} ) \phi (\vec{y} ) \rangle $ decay with their distance 
$ | \vec{x} - \vec{y} | $. 
The connected correlation function is defined as
\begin{equation}
G_{s}(\vec{r},\vec{r}^{\hskip .1truecm \prime})=\langle
\phi (\vec{r}) \psi (\vec{r})
\phi (\vec{r}^{\hskip .1truecm \prime}) \psi (\vec{r}^{\hskip .1truecm \prime})
 \rangle
\end{equation}
The subscript $s$ represents the dependence on the random field sample. 
We can also define $G^{(k)}(\vec{r}-\vec{r}^{\hskip .1truecm \prime})$
the average over the samples of the $k$th moment
of $G_{s}(\vec{r},\vec{r}^{\hskip .1truecm \prime})$
(averaging over the random fields restores
translation invariance), $G^{(k)}(\vec{r}-\vec{r}^{\hskip .1truecm \prime}) = 
\overline { (G_{s}(\vec{r},\vec{r}^{\hskip .1truecm \prime}))^k } $. 
These correlations $G^{(k)}(\vec{r}-\vec{r}^{\hskip .1truecm \prime})$ 
can be simply defined if, as usually,  
we introduce $n$ replicas 
(at the end $ n \to 0 $) to average over the random fields. 
In the high temperature phase and for large $ |\vec{r}-\vec{r}^{\hskip .1truecm \prime } | $,
$ G^{(k)}(\vec{r}-\vec{r}^{\hskip .1truecm \prime}) \sim { \exp -( \mu (k)
 |\vec{r} - \vec{r}^{\hskip .1truecm \prime} | ) \over
|\vec{r} - \vec{r}^{\hskip .1truecm \prime} |^{\alpha (k) } } $
If we sum over the ``transverse" $y$ and $z$ components of the vectors
$\vec{r}$ and $\vec{r}^{\hskip .1truecm \prime}$,
$ G^{'(k)}(r_{x}-r^{'}_{x}) =      
\int_{r_{y},r_{z},r^{'}_{y},r^{'}_{z}}
 G^{(k)}(\vec{r}-\vec{r}^{\hskip .1truecm \prime}) $, then 
$ G^{'(k)}(r_{x}-r^{'}_{x}) \sim { \exp -( \mu (k) |r_{x} - r^{'}_{x} | ) \over 
|r_{x} - r^{'}_{x} |^{\alpha^{'} (k) } } $
 i.e. $G^{'(k)}(r_{x}-r^{'}_{x}) $
and $ G^{(k)}(\vec{r}-\vec{r}^{\hskip .1truecm \prime}) $ have the same
exponential behaviour,
only the power prefactor is different. 
If we define
\begin{equation}
Q^{\alpha} (x) = { 1 \over L^2 } \sum_{y,z} \phi^{\alpha} (x,y,z ) 
\psi^{\alpha} (x,y,z ) 
\end{equation}
where $\phi$ and  $\psi$ are the fields corresponding to two identical 
copies of the system, and the ``composite" operators 
$A^{\alpha^{1} \alpha^{2} \cdots \alpha^{k} } (x) = Q^{\alpha^1} (x)  
Q^{\alpha^2} (x) 
\cdots  Q^{\alpha^k} (x) $, where $\alpha^{1}, \alpha^{2} \cdots \alpha^{k} $ 
are $k$ replica indices all different from each other,
\begin{equation} \label{gp}
G^{'(k)}(x-y) \ = \ \langle A^{\alpha^{1} \alpha^{2} \cdots \alpha^{k} } (x) 
A^{\alpha^{1} \alpha^{2} \cdots \alpha^{k} } (y) \rangle
\end{equation}
We can insert a complete set of states $r$ in equation \ref{gp}
\begin{equation} 
G^{'(k)}(x-y) = \sum_{r} \langle A^{\alpha^{1} \alpha^{2} \cdots \alpha^{k} } (x)
| r \rangle \langle r |
A^{\alpha^{1} \alpha^{2} \cdots \alpha^{k} } (y) \rangle
\end{equation}
Let's call $r_{0} (k) $ the lowest mass state giving a non zero contribution to the previous 
sum and $ \mu (k) $ its masse. 
For large $ |x-y| $, $ G^{'(k)}(x-y) \sim \exp -( \mu (k) x) $.  
In perturbative field theory these lowest masse states $r_{0} (k) $ 
are those created by 
the $2k$ fields $ \phi^{\alpha^{1}}, \ \psi^{\alpha^{1}}, \ \cdots , \ 
\phi^{\alpha^{k}}, \ \psi^{\alpha^{k}}$, i.e., $ \mu (k) =  2 k \mu $, 
where $ \mu $ is the masse of the field $\phi $. 
If we find $ \mu (k) \ne 2 k \mu $ perturbation theory brakes down. 

If universality is valid along the transition line, down to zero 
temperature, as it is usually assumed, the previous arguments are also valid 
at zero temperature. First remark that for 
$J < J_c $, and for $ x > > 1 $, $q_{s}(x) = 1 $, so the ``connected" part 
of $q_{s}(x) $ is $q_{s,c}(x) = q_{s}(x) -1 = -2 d_{s}(x) $. It follows 
from the previous that 
the long distance behaviour of $d^{(k)} $ and $ G^{'(k)}(x-y) $ are the same. 
We found in our simulations that, near the critical region, $ m(k) \ne k m(1) $,  
 i.e. there 
exists a new state $r_0$, coupled to all the composite operators 
$A^{\alpha^{1} \alpha^{2} \cdots \alpha^{k} } $ (we have only measured 
the correlation functions for $k=1,2,3$). 
This means that there exists a bound state and this explains why 
perturbation theory breaks down in the RFIM. 

A preliminary version of this work was presented by one of the 
authors (NS) at the Helsinki Workshop Disordered systems at low temperatures 
 and their topological properties (January 2002).

This work was partially supported by the SPHINX programme of the European Science 
Foundation.

\bibliographystyle{prsty}
\bibliography{../../../Bib/references}

 \end{document}